# Modelling approaches to capture role of gelatinization in texture changes during thermal processing of food

**Ankita Sinha[1], Atul Bhargav[1,*]**

[1]Department of Mechanical Engineering
Indian Institute of Technology Gandhinagar
Palaj, Gandhinagar, GJ 382355
India

[*]**Corresponding author**
Tel: +91 792 395 2419, +91 814 030 7813
Email: atul.bhargav@iitgn.ac.in  (Atul Bhargav)

# Modelling approaches to capture role of gelatinization in texture changes during thermal processing of food

**Abstract**

While processing at elevated temperatures, starchy food products undergo gelatinization, which leads to softening-related changes in textural characteristics. Study of role of gelatinization in texture development is limited; specifically, ignoring gelatinization physics can lead to erroneous prediction of Young's modulus. In a recent work, we demonstrated a texture model that predicts local and effective Young's moduli as functions of moisture content. While this model tracks experiments closely for drying, it deviates significantly from experiments for frying. In this paper, three different techniques for incorporating starch gelatinization are used to enhance the texture model, and the suitability of each technique is discussed. These improved models can capture the initial gelatinization-induced softening and are in much better agreement with experiments. We expect that this work could contribute to better understanding and predictive capabilities of texture development.

## 1. Introduction

Starch gelatinization is a physio-chemical process that involves disaggregation of starch granules in the presence of heat and water (Zanoni et al., 1995). When heated in an aqueous medium, starch molecules undergo swelling, which becomes irreversible beyond a certain temperature, and then the diffusion of material from the starch granules into water begins. This temperature is called the gelatinization temperature, and marks the onset of the gelatinization process (Lund and Lorenz, 1984). Marchant and Blanshard (1978) explained gelatinization as a three-step mechanism that involves diffusion of water in the molecules, transition of helical coils and swelling due to crystal disintegration. Various experimental techniques used for analysing starch gelatinization include differential scanning calorimetry (DSC) (Stevens and Elton, 1971), X-ray diffraction (Lugay and Juliano, 1965; Owusu-Ansah et al., 1982; Varriano-Marston et al., 1980), birefringence end point method (Watson, 1964) etc. DSC is the most common method, and has been used extensively for analysing starch gelatinization in different types of starchy food materials (Altay and Gunasekaran, 2006; Larsson and Eliasson, 1991; Liu and Thompson, 1998; Schirmer et al., 2011; Wolters and Cone, 1992; Zanoni et al., 1991). Kinetic models have been proposed on the basis of experiments that approximate starch gelatinization as first order reactions (Lund and Wirakartakusumah, 1984; Zanoni et al., 1995). Alhough first order reaction models have been extensively used for modelling gelatinization, some researchers have showed that it is a multi-step process (Pielichowski et al., 1998; Svensson and Eliasson, 1995). At lower temperatures, the complete gelatinization of starch does not take place (Pravisani et al., 1985; Wirakartakusumah, 1982). To account for the maximum degree of starch gelatinization at lower temperature and prolonged heating, a modified first order reaction model has been developed (Pérez-Santos et al., 2016). Besides kinetic models, a three parameter Weibull model has also been used to model starch gelatinization and digestibility as a function of temperature and water activity in plantain flour (Toro et al., 2015). In a recent study, starch gelatinization and *in vitro* digestibility of plantain during steeping is predicted using Weibull model, and is coupled with heat and mass transfer using finite element based modelling (Toro et al., 2018). The major challenge with kinetic and logistic modelling is the estimation of kinetic and logistic parameters and their replicability.

Gelatinization-induced softening (also called initial softening or thermal softening) affects texture development (Alvarez et al., 2001; Alvarez and Canet, 2000; Andersson et al., 1994; Huang and Bourne, 1983; Taherian, 1996). Potato frying can therefore be considered as a two stage process: the first stage corresponds to the softening phase, and crust hardening occurs in the second (Pedreschi et al., 2004, 2001; Pedreschi and Moyano, 2005)

Kinetic models for texture prediction that account for this softening are limited. Verlinden et al. (1995) proposed a two-compartment model to account for textural changes due to gelatinization. The total texture is defined as the sum of two terms: the first term corresponds to texture development due to the cell wall, and the second term accounts for starch gelatinization. In an alternative texture modelling approach proposed for frying process, the initial softening period during which significant texture loss occurs is neglected (Nourian and Ramaswamy, 2003). Study of kinetic of starch gelatinization in reference to texture development is significant as it can lead to improved and more accurate texture prediction model.

This paper investigates three different methodologies to incorporate softening due to starch gelatinization in the texture model that is proposed in [cite paper 1]. A comparison is shown for the three methodologies and their suitability is discussed.

## 2. Starch Gelatinization modelling

Starch gelatinization can be approximated by a first order kinetic system with a parameter $S$ representing the fraction of un-gelatinized starch (Verlinden et al., 1995).

at time $t = 0$ $\qquad S = S_O \qquad$ (1)

Following first order reaction:

$$\frac{dS}{dt} = -k_g S \qquad (2)$$

where
$$k_g = k_g^{ref} exp\left(\frac{-E_{ag}}{R}\left(\frac{1}{T} - \frac{1}{T_{ref}}\right)\right) \qquad (3)$$

Gelatinization begins only when system reaches a threshold temperature called the gelatinization temperature $T_g$ (Lund and Lorenz, 1984). In a thermal process, let $t_g$ be the time taken by the system to reach gelatinization temperature. Equation (2), when integrated with initial condition $t = t_g$ gives:

$$S = S_o exp(-k_g(t - t_g)) \qquad (4)$$

Alternatively, the process can be defined in terms of fraction of gelatinised starch ($\alpha$) present at any instant (Pielichowski et al., 1998; Toro et al., 2015; Zanoni et al., 1995, 1991).

$$\alpha = 1 - S \qquad (5)$$

Thus, equation (4) can be written as:

$$\alpha = 1 - (S_o \exp(-k_g(t - t_g))) \qquad (6)$$

Verlinden et al. (1995) demonstrated that gelatinization occurs over a relatively short period of time. This agrees well with initial softening that occurs during frying for a short time. Thus, by incorporating the gelatinization mechanism in texture models, initial softening phase could be captured, leading to more realistic prediction of textural characteristics.

## 3. A texture model neglecting initial softening [cite paper 1]

[cite paper1] proposed texture models to predict variations in local and effective Young's modulus as a function of moisture content (and thus treatment time). Equations are as follows:

$$E = E_o \, exp\left(\left[ln\frac{E_{cr}}{E_o}\right]\left[\frac{M_o - M}{M_o - M_{cr}}\right]\right) \qquad (7)$$

$$E_{eff} = \sqrt[n]{\prod_{k=1}^{n} E_k} \qquad (8)$$

This model is validated with drying and frying experiments on potato samples at $T = 115°C$ and $T = 170°C$ respectively. Variation in Young's modulus with respect to moisture content for these processes are shown in Figure 1 and Figure 2 respectively. Figure 3 shows the variation in Young's modulus with time during deep frying. Model results agree well with experiments for drying, whereas a relatively large variation from experimental results is observed for deep frying. We attribute this departure to the initial softening process that occurs above the gelatinization temperature.

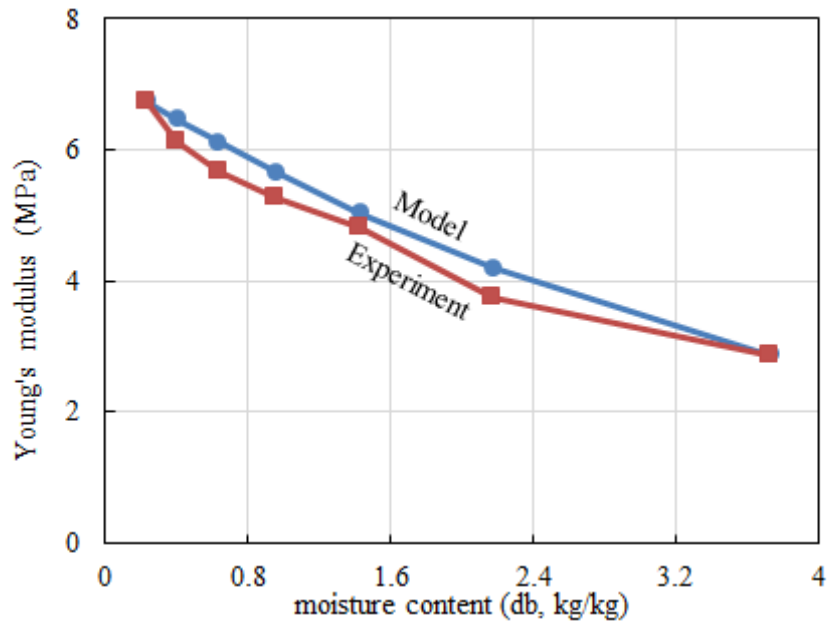

Figure 1: Young's modulus increases with decrease in moisture content during drying. Experiments are performed to measure local Young's modulus using 14 X 14 X 2 mm³ potato samples dried at = 115°C .

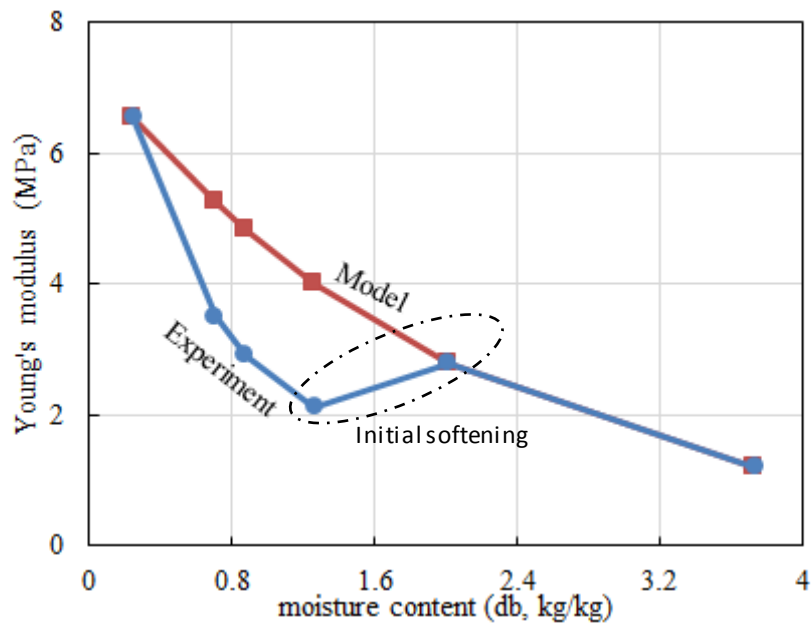

Figure 2: Young's modulus does not increase monotonously with decrease in moisture content during deep frying, possibly due to the so-called initial-softening effect, caused by starch gelatinization. Experiments are performed in corn starch oil using 14 X 14 X 2 mm³ potato samples at T = 170°C.

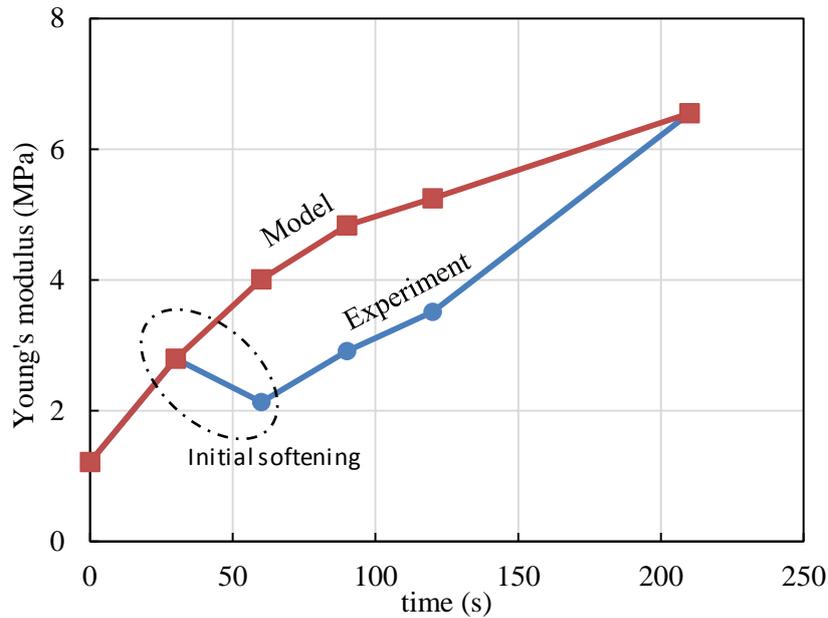

Figure 3: Variation in Young's modulus with time during deep frying. Initial softening effect occurs over a short span of time.

Importantly, the initial softening is not observed during drying at carried out at $T = 115°C$ Figure 1 because for drying temperature above boiling point of water, continuous and rapid loss of near-surface moisture takes place from the food sample. This leads to crust hardening which causes increase in Young's modulus. Thus, during drying carried out at high temperature, crust hardening dominates over initial softening, leading to overall increase in Young's modulus.

## 4. Incorporating initial softening behaviour in texture model: proposed methodologies

The following three approaches have been developed to include the initial softening effect in existing models for the deep-frying process:

a) Model with modified input parameters
b) A two-compartment additive model
c) A two-compartment multiplicative model

Each of the methodology is explained in detail as follows:

### 4.1 Model with modified the input parameters

The original texture model is valid for relatively short durations toward the beginning and the end of deep frying (Figure 2 and Figure 3), deviating significantly from experimental values during intermediate times; this can be attributed to the lack of gelatinization physics in the model. By ignoring the region before the gelatinization (points A and B) and considering point C as the model's new initial condition, the model can now predict the trajectory C-D-E-F with sufficient accuracy (Figure 4, Figure 5 and Figure 6).

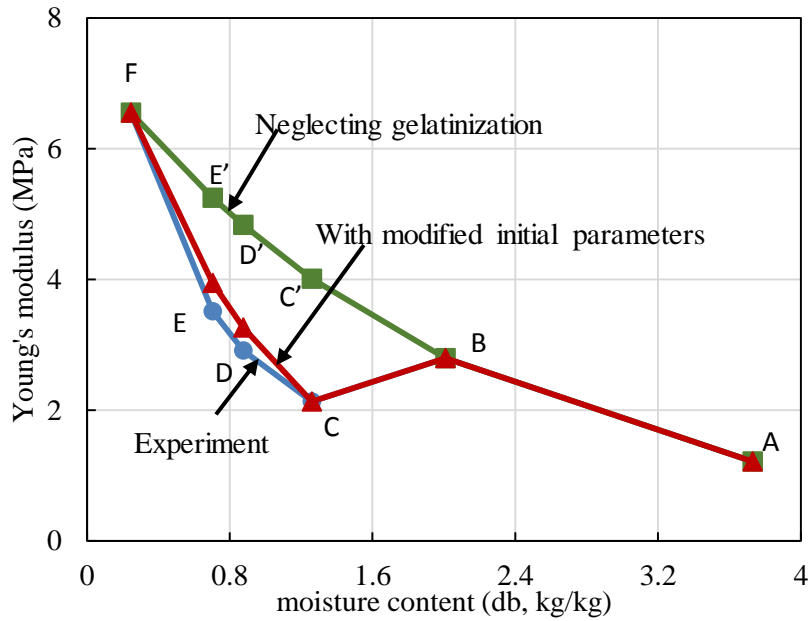

Figure 4: Variation in Young's modulus with moisture content. By modifying input parameters, the model captures texture development more efficiently.

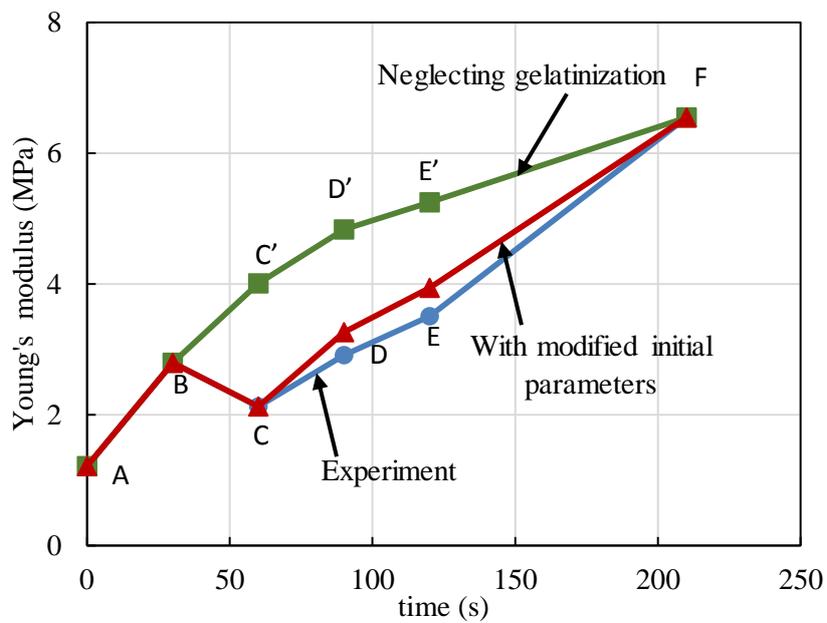

Figure 5: Variation in Young's modulus with time: Model with modified input parameters vs. model neglecting gelatinization.

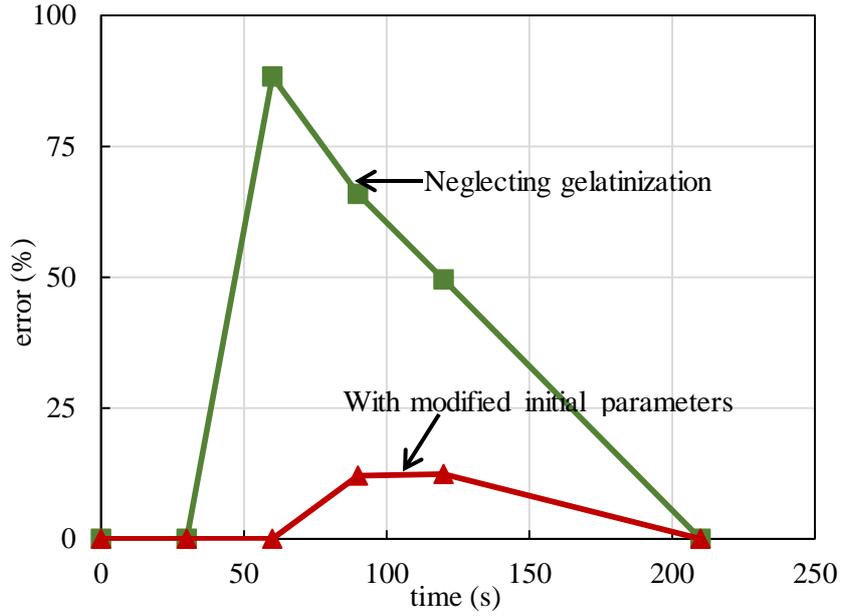

Figure 6: Error analysis shows reduction in error by using modified input parameters.

The obvious drawback of this model is that properties at the end of gelatinization have to be known *a priori*, which is not practical in most situations.

### 4.2 A two-compartment additive model

One of the initial works that emphasise the need of incorporating gelatinization in texture model is compartmental texture model (Verlinden et al., 1995). In this model, the rupture force ($F$) was used as a measure of texture, and overall texture of the food sample is considered to be a linear combination of two terms. The first term corresponds to texture development due to ungelatinized starch and second due to all other factors, mainly cell wall structure.

$$F = F_s + F_c \qquad (9)$$

Further,

$$F_s(t) = S(t)F_{so} \qquad (10)$$

where $S(t)$ is the fraction of ungelatinized starch at time $t$, and $F_s$ is texture associated with it ($F_s(t=0) = F_{so}$). $F_c$ is the texture developed due to cell wall structure and other relevant factors. A first order kinetic model was used to determine $S(t)$ and $F_c(t)$ and model parameters were determined by fitting model to the experimental data.

Assuming a linear combination (equation (9)), and considering Young's modulus as the measure of texture,

$$E = E_s + E_c \qquad (11)$$

where

$$E_s = SE_{so} \tag{12}$$

Further, the following hypotheses were made:

a) $S = \begin{cases} 1, & t < t_g \\ 1 > S \geq 0, & t \geq t_g \end{cases}$ \hfill (13)

b) $E_s = \begin{cases} E_{so}, & t < t_g \\ SE_{so}, & t \geq t_g \end{cases}$ \hfill (14)

c) $\frac{\partial E_c}{\partial t} > 0, \ \forall\, t > 0$ \hfill (15)

d) $E(t) = \begin{cases} E_{co} + E_{so}, & t = 0 \\ E_c(t) + E_{so}, & 0 < t < t_g \\ E_c(t) + S(t)E_{so}, & t \geq t_g \end{cases}$ \hfill (16)

*Proposed modification in the texture model*

In the proposed model (equation (7)), $E_o$ is the initial Young's modulus. It is measured experimentally, and is therefore a combination of initial values of $E_c$ and $E_s$ i.e.

$$E_o = E_{co} + E_{so} \tag{17}$$

Further, for prediction of intermediate values no consideration is made to account for variation in $E_s$. Thus, values predicted by the proposed texture model (equation (7)) can be written as:

$$E_{pred}(t) = E_c(t) + E_{so} \tag{18}$$

However, a better model would be:

$$E_{corr}(t) = E_c(t) + S(t)E_{so} \tag{19}$$

Solving equation (18) and equation (19), we get:

$$E_{corr}(t) = E_{pred}(t) + (1 - S(t))E_{so} \tag{20}$$

Here $E_{pred}$ is $E$ value predicted by equation (7) and $E_{corr}$ is the corrected $E$ value when gelatinization is taken into account.

*S(t) estimation:*
From equation (4)

$$S(t) = S_o \exp\left(-k_g(t - t_g)\right) \tag{21}$$

where $k_g$ is a Arrhenius-type rate constant and has the expression given in equation (22).

$$k_g = k_{go}\exp(-E_a/RT) \tag{22}$$

$k_g$ is a function of temperature. However, for modelling local Young's modulus, $k_g$ can be assumed constant spatially as well as temporally by assuming constant temperature during gelatinization i.e. $\frac{\partial T}{\partial t} \approx \frac{\partial T}{\partial x} \approx \frac{\partial T}{\partial y} \approx \frac{\partial T}{\partial z} \approx 0$ during gelatinization

The above claim can be justified by the following reasons:

a) Due to small sample size (14 X 14 X 2 mm³), temperature can be assumed near uniform in the sample and thus can be treated constant over whole domain at any instant ($\frac{\partial T}{\partial x} \approx \frac{\partial T}{\partial y} \approx \frac{\partial T}{\partial z} \approx 0$).

b) Since gelatinization is a short-term phenomena (occurs over a span of few seconds), thus temporal variation can be neglected ($\frac{\partial T}{\partial t} \approx 0$).

Variation in $S$ can thus be is estimated by curve fitting in softening region and $k_g$ is predicted by the trial and error method.

$E_{so}$ estimation:
There is no reported way to calculate $E_{so}$. As derived earlier in equation (16):

$$E_{corr}(t) = E_{pred}(t) + (1 - S(t))E_{so} \tag{23}$$

Assuming that the difference in predicted and experimental values is only due to ignoring the gelatinization term:

$$E_{so} = \frac{\sum_i \frac{E_{exp}(i) - E_{pred}(i)}{1 - S(i)}}{n} \tag{24}$$

($i$: terms where prediction failed; n=total no. of $i$)

For the terms where $E_{pred}$ failed; $E_{pred}$ is replaced by $E_{corr}$ from equation (19); with $E_{so}$ substituted from equation (24).

Results: Comparison of additive model with the initial model that neglects gelatinization and experiments is shown in Figure 7 and Figure 8. The proposed additive model shows good agreement with the experimental results. Error analysis shows that absolute error for additive model is less than 5% at all times (Figure 9).

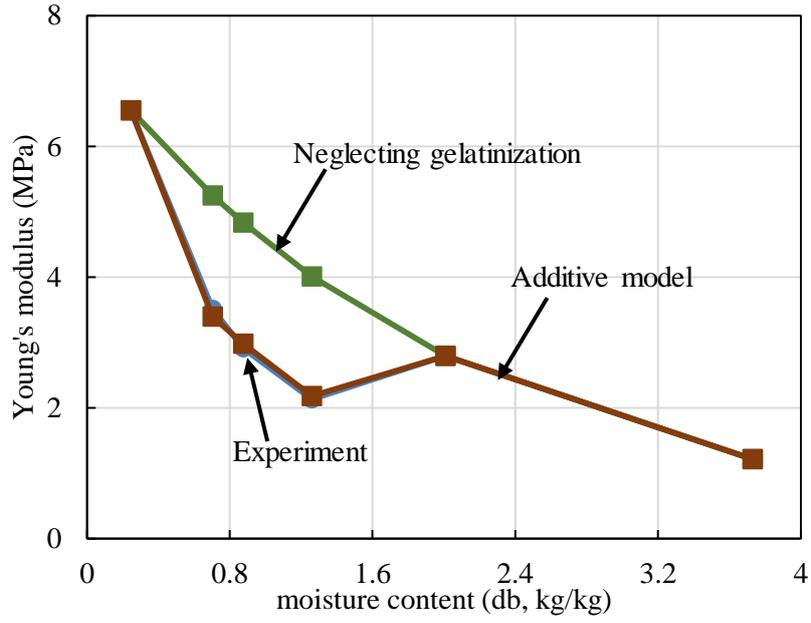

Figure 7: Variation in Young's modulus with moisture content. Additive model fits well at all points.

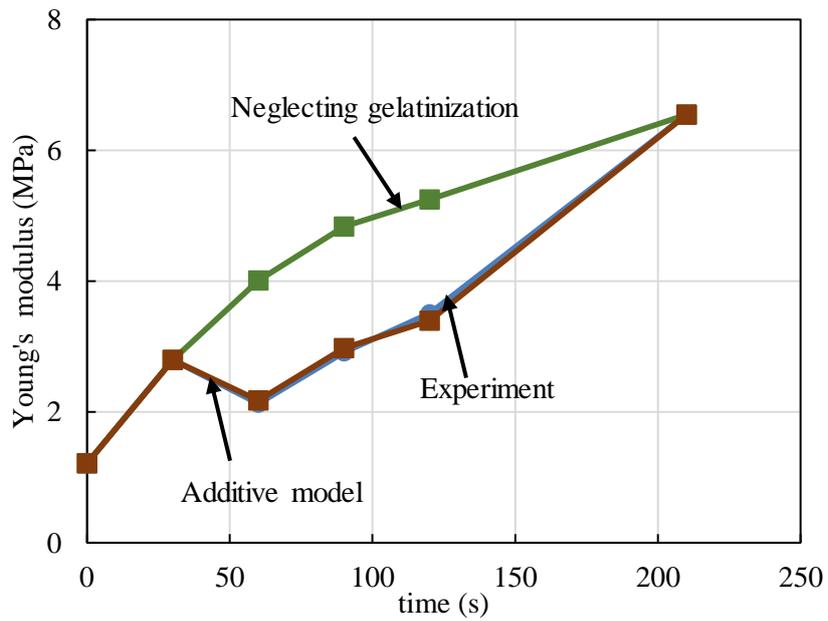

Figure 8: Variation in Young's modulus with time: Additive model vs. model neglecting gelatinization.

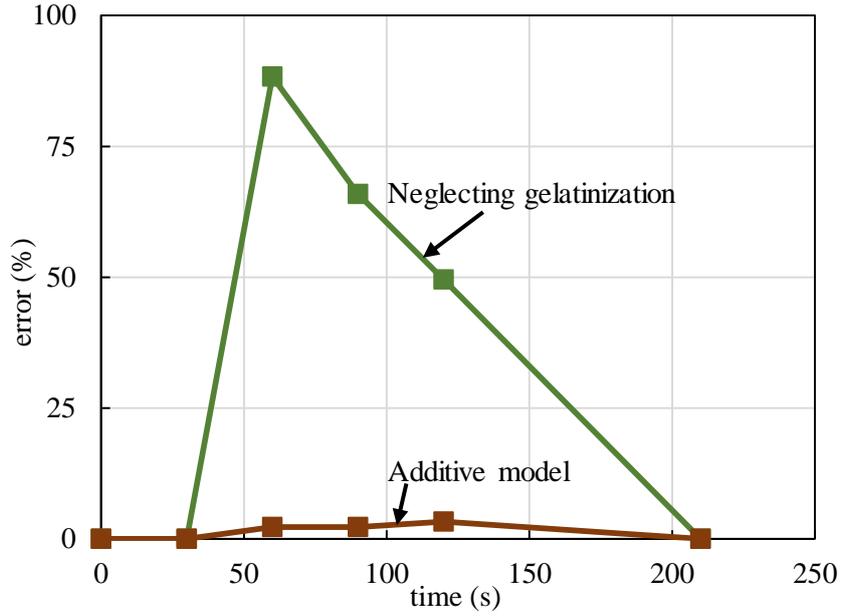

Figure 9: Using additive model, the maximum obtained error reduces to less than 5%.

### 4.3 A two-compartment multiplicative model

According to Verlinden et al. (1995), texture is a linear combination of starch induced texture and cell wall induced texture; equation (19) is based primarily on this hypothesis. However, this assumption may not be accurate, as explained below.

By equation (18)

$$E_{pred}(t) = E_c(t) + E_{so} \qquad (25)$$

Figure 7 and Figure 8 show that the prediction model fits well toward the end of the process, which is not consistent with equation (25), since $E_s \neq E_{So}$. Also, we have shown previously that Young' Modulus varies exponentially with moisture content [cite paper 1]. Thus, introducing gelatinization in the proposed model in the form of a multiplicative exponential term could be a better methodology as the modified model could now be valid at all points in the domain.

Let

$$M^* = \frac{M_0 - M}{M_0 - M_{cr}} \qquad (26)$$

$$\alpha = 1 - S \qquad (27)$$

Young's modulus increases with increase in $M^*$ as well as decrease in $\alpha$, the modified prediction model is of the form:

$$E = A\exp(BM^* - C\alpha) \quad (28)$$

where $A$, $B$ and $C$ are constants.

At $\alpha = 0, M^* = 0$

$$E = E_0 \quad (29)$$

At $\alpha = 1, M^* = 1$

$$E = E_{cr} \quad (30)$$

Solving equations (28) – (30), we get

$$E = E_0 \exp\left(\left(\ln\frac{E_{cr}}{E_o}\right)M^* - C(\alpha - M^*)\right) \quad (31)$$

$C$ is obtained here by trial method. This model is now valid at all points during the process.

Results: Figure 10 and Figure 11 show comparison of multiplicative model with experimental results (and the original model that neglects gelatinization). The maximum error obtained in this multiplicative model is about 12% (Figure 12).

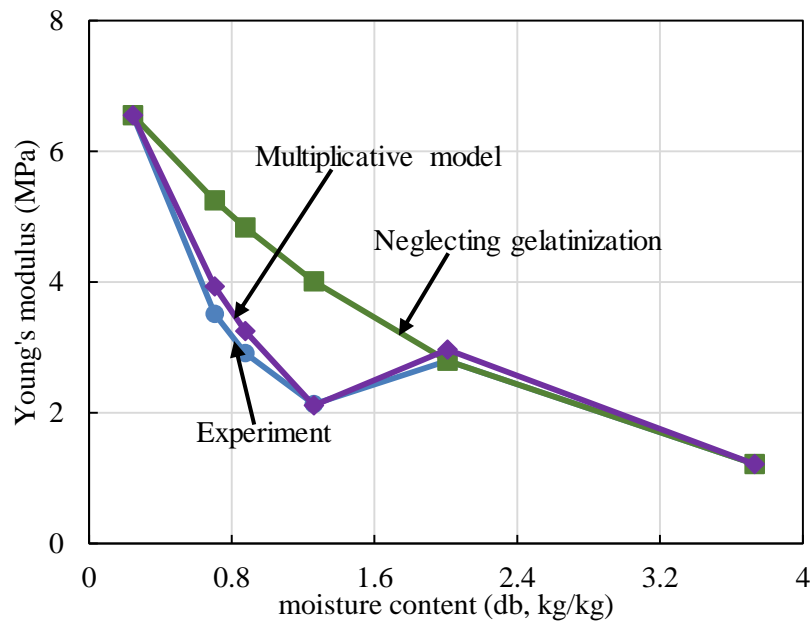

Figure 10: Variation in Young's modulus with moisture content. Multiplicative model captures the physics well in all regions.

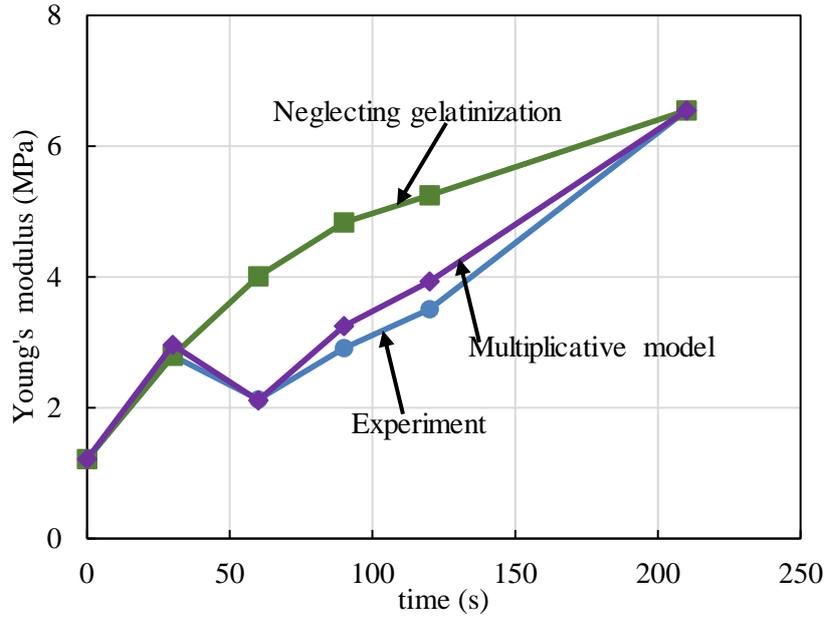

Figure 11: Variation in Young's modulus with time: Multiplicative model vs. model neglecting gelatinization.

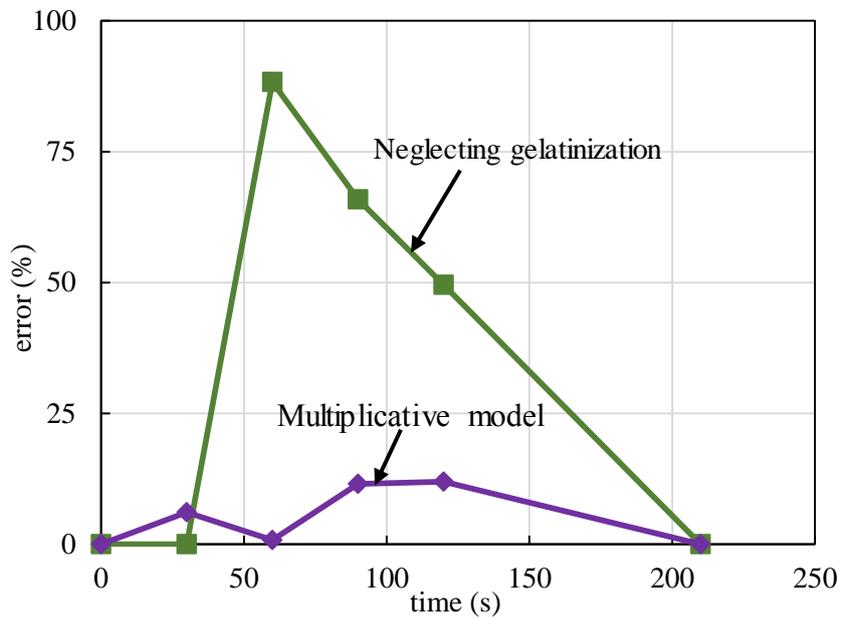

Figure 12: Error analysis shows that maximum observed error reduces to around 12% using multiplicative model.

## 5. Comparison of the models

The results obtained by all three proposed methodologies are in good agreement with experiments. A comparison is shown in Figure 13 and Figure 14. Figure 15 shows the % error obtained in case of each of the methodology. The minimum error is obtained in case of model 2 (<5%). However, a more viable justification for the process is achieved by model 3. The deviation from experiments could also be due the error in estimating fitting parameters, or due to experimental variation.

To further validate the models, they are also tested on experimental data from a different source, where convection drying at relatively lower temperature is performed (Yang, 2001). In this case, due to low drying temperature, the dehydration and crust hardening is not rapid enough to neglect initial softening. Therefore, prediction model needs to take into account the softening mechanism. This is shown in Figure 16. Figure 17 shows the error analysis for the models. Prediction model that does not account for gelatinization has relatively poor match with experiments.

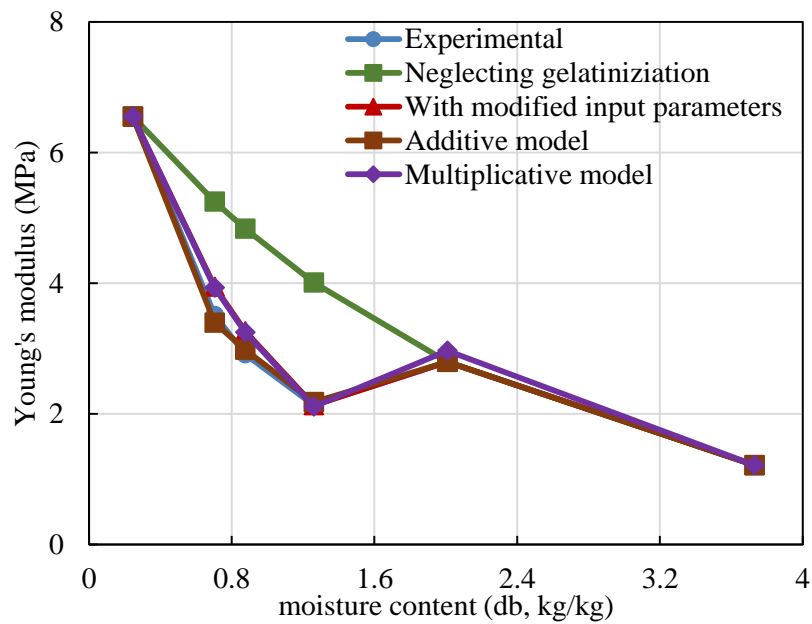

Figure 13: Variation in Young's modulus with moisture content. Comparison of models show that model with modified input parameters and multiplicative model show similar results. Additive model fits to experiments better than other models.

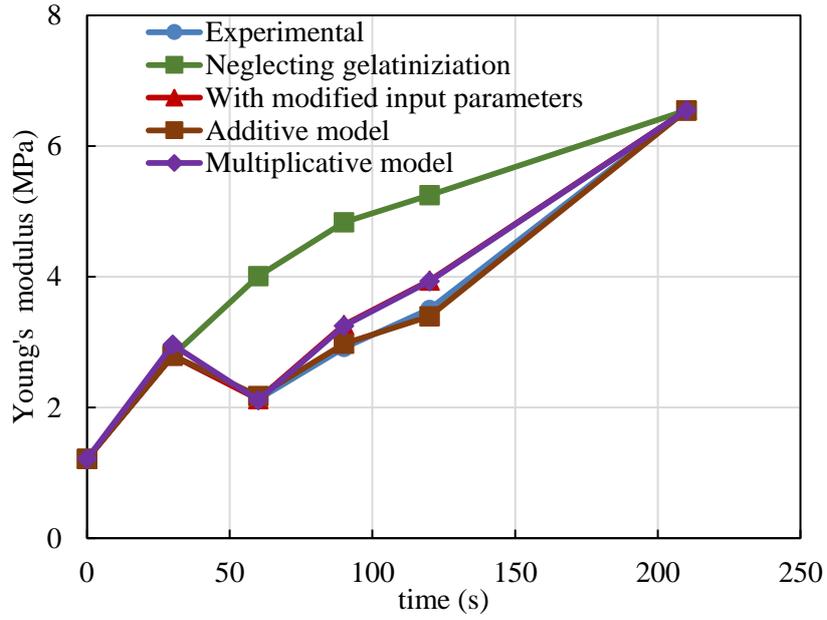

Figure 14: Variation in Young's modulus with time: Comparison of models.

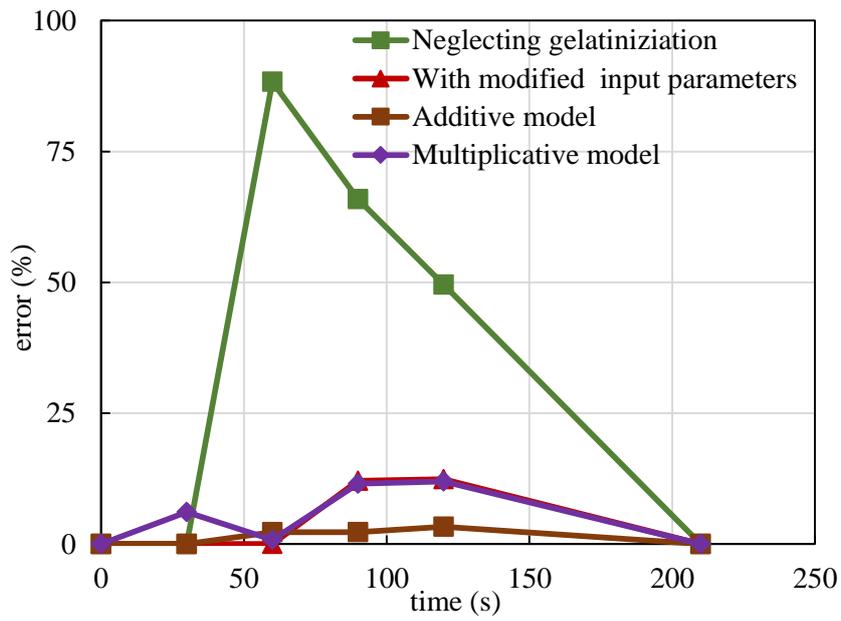

Figure 15: Error analysis: Comparison of models.

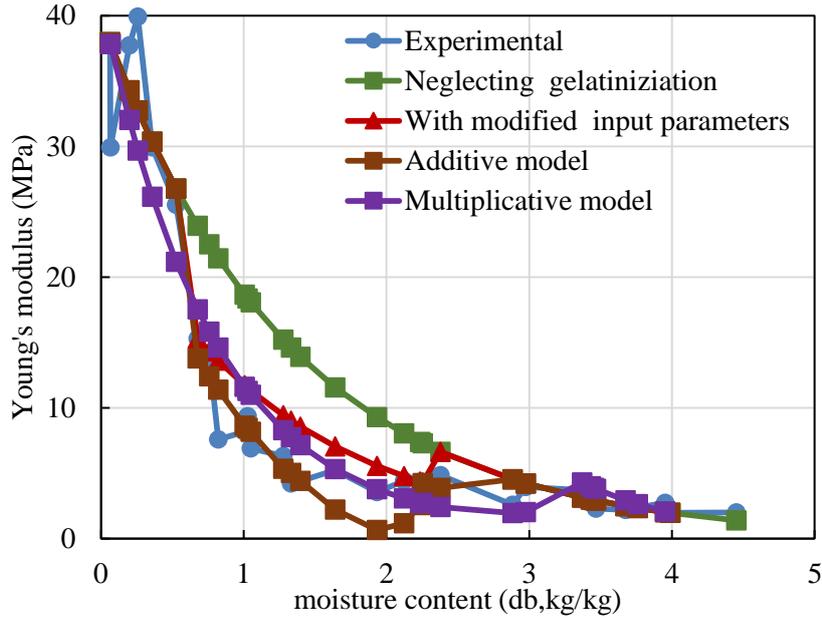

Figure 16: Variation in Young's modulus with moisture content: experimental data taken from Yang(2001).

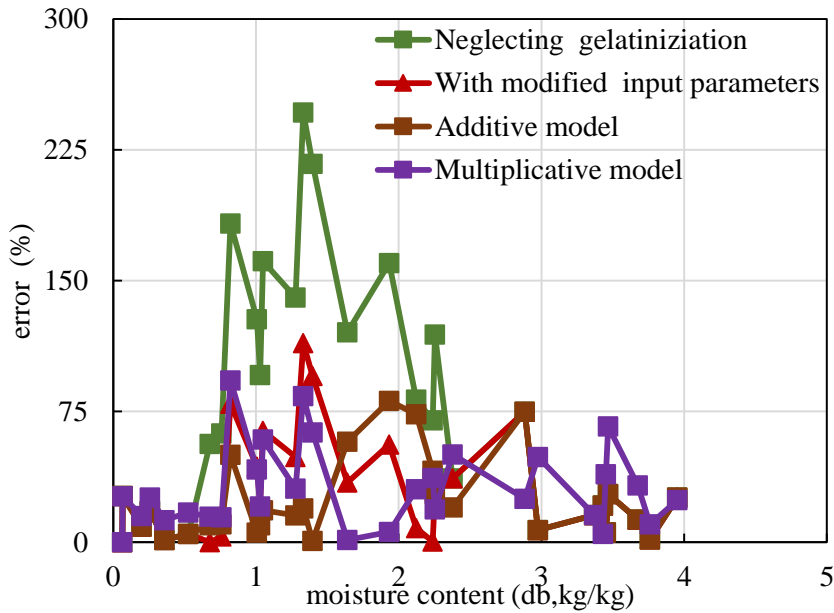

Figure 17: Error analysis: error reduces using models that incorporate gelatinization.

## 6. Conclusions

Role of starch gelatinization in texture development of starchy food products is emphasized. Three approaches to account for gelatinization in texture model are presented. Though all the three approaches predict the scenario well, third model is suggested as the best choice as it is defined well in all regions and is more justifiable. The proposed models are not standalone as gelatinization prediction is done on the basis of experimental softening data. This offers future scope for improvement by developing and implementing better modelling techniques for gelatinization mechanism.

## Nomenclature

| | |
|---|---|
| $E$ | Young's modulus (MPa) |
| $M$ | moisture content (dry basis, kg/kg) |
| $S$ | fraction of ungelatinized starch (unit less) |
| $T$ | Temperature (K) |
| $F$ | rupture force (N) |
| $R$ | Universal gas constant (J/mol-K) |
| $E_{ag}$ | activation energy for gelatinization (J/mol) |
| $M^*$ | non-dimensional moisture content (unit less) |
| $t$ | time (s) |
| $\alpha$ | fraction of gelatinized starch (unit less) |
| $k_g$ | gelatinization rate coefficient (1/s) |

subscripts

| | |
|---|---|
| $ref$ | reference |
| o | initial |
| $cr$ | critical |
| $av$ | average |
| $k$ | element index |
| $n$ | Total number of elements |
| $c$ | Cell wall |
| $s$ | starch |
| $g$ | gelatinization |
| $eff$ | effective |
| $exp$ | experimental value |
| $pred$ | predicted value |
| $corr$ | corrected value |

## Funding

Ankita Sinha receives stipend support from the Ministry of Human Resource Development (MHRD), Govt. of India. This research did not receive any specific grant from funding agencies in the commercial, public or not-for-profit sectors.

## Acknowledgments

The authors are thankful to College of Food Processing Technology and Bio-Energy at Anand, Gujarat (India) for providing access to their experimental facilities for this research.